\documentclass[aps,floatfix,showpacs,superscriptaddress,prd]{revtex4}

\usepackage[dvips]{color,graphicx}
\usepackage{epsfig}
\usepackage{bm}
\usepackage{pstricks}
\usepackage{graphicx}
\usepackage{dcolumn}
\usepackage{bm}
\usepackage{natbib}

\begin{document}

 
\title{
{\tiny{DESY 11-263, DO-TH 12/02, SFB/CPP-12-01, LPN12-002  \hfill}}\\
Unfolding of target mass contributions from inclusive proton structure  
function data} 



\author{M.~E.~Christy}
\affiliation{Hampton University, Hampton, VA 23668, USA}
\email{christy@jlab.org} 

\author{J. Bl\"{u}mlein}
\affiliation{Deutsches Elektronen‐Synchrotron, DESY, Platanenallee 6, D‐15738 Zeuthen, Germany } 

\author{H. B\"{o}ttcher}
\affiliation{Deutsches Elektronen‐Synchrotron, DESY, Platanenallee 6, D‐15738 Zeuthen, Germany }






\date{\today}

\begin{abstract}

We report on the extraction of the target mass contributions to the unpolarized proton 
structure functions by applying an unfolding procedure to the available world data from 
charged lepton scattering.  The method employed is complementary to recent and future 
parton distribution function fits including target mass contributions and the results 
obtained can be utilized to further study perturbative QCD at large Bjorken $x$ and 
small $Q^2$.  Global fits are performed to both the available $F_2^p$ and the separated 
$F_L^p$ ($F_1^p$) data which yield excellent descriptions and provides parameterizations 
of the extracted target mass contributions.

\end{abstract}


\maketitle

\bibliographystyle{apsrev}

\section{Introduction}

Since the inception of QCD as {\it the} theory of the strong interaction, perturbative QCD (pQDC) 
has been spectacularly successful in explaining a host of scattering and annihilation processes 
involving hadrons.  Among these is the $Q^2$ dependence of the proton $F_2$ structure function 
measured in charged lepton deeply inelastic scattering (DIS).  However, the separation of the 
{\it pure} leading twist pQCD $Q^2$ dependence from possible higher twist (HT) and so called target mass 
contributions (TMCs) is 
required for any quantitative study of pQCD at moderate to large values of Bjorken $x$ and low $Q^2$.  
The 
large body of existing data on $F_2^p$ covers a significant range in $Q^2$ and $x$, and is a 
critical input for pQCD based fits which attempt to extract the parton distribution functions 
(PDFs).  While the contributions from TMCs are now being included in some PDF 
fits, cf. 
e.g.~\cite{Alekhin:2000ch,Blumlein:2004ip,Blumlein:2006be,alehkin,cj1}. 
Historically 
the data which were expected to have significant 
TM or HT contributions were excluded in standard analyses 
from the fits via kinematic cuts on the invariant hadron mass, $W$.  Typically, this has removed 
nearly all the precision data from SLAC at moderate to large $x$ from the fits, significantly 
reducing the $Q^2$ lever arm on the remaining data, and resulting in the rather large error bands 
quoted from these fits for $x > 0.5$.   

In this paper we present a fit to the world charged lepton scattering data on the proton $F_2$ 
and $F_L$ ($F_1$) structure functions with the aim of separating the TMCs from the massless 
limit structure functions via an unfolding procedure. The method employed is complementary to PDF 
fits including the twist-2 target mass contributions \cite{gp}, but has the advantage that it does 
not rely on a finite expansion in the strong coupling constant $\alpha_s$.  This avoids the issue 
of disentangling higher order QCD corrections from the TMCs. The extracted distributions still contain 
the higher twist terms. 

\section{Data Sets}

The world charged lepton scattering data sets utilized for the $F_2$ fit are listed 
in Table~\ref{datasets-f2} and include data from the BCDMS~\cite{f2rbcdms}, 
and  NMC~\cite{f2rnmc} collaborations, the reanalyzed SLAC data from Whitlow 
$et$ $al.$~\cite{f2whitlow}, and the $e^+$ and $e^-$ data from the H1~\cite{f2h1e+,f2h1e-} and 
ZEUS~\cite{f2zeuse-,f2zeuse+} collaborations; the latter two help to constrain the fit at both 
low $x$ and large $Q^2$.  To help constrain the low $Q^2$ region, recent data from 
Jefferson Lab Hall C experiments E99-118~\cite{f2e99118} and E94-110~\cite{e94110} have been 
utilized.    

These experiments have additionally performed limited longitudinal and transverse (L/T) separations 
to extract the information on the transverse structure function, $F_1$, the longitudinal structure 
function, $F_L$, and the ratio $R = F_L / 2xF_1$.  For the $F_L$ fit we include {\it only} the data 
extracted via L/T separations in order to minimize 
the correlation between the $F_2$ and $F_L$ data sets.  For the JLab experiments, only the L/T 
separated values for $F_2$ were utilized.  
It is unfortunate that more experiments have not made readily available the complete set of 
separated structure functions extracted from L/T separations.  Instead, 
what can generally be found is a table of the extracted $R = \sigma_L / \sigma_T = F_L/2xF_1$ 
values, but without the consistently extracted values for $F_2$ (or $F_1$).  In these cases the 
'data' on $F_L$ have been constructed utilizing the available data on $R$ and a parameterization 
of the $F_2$ structure function.  The longitudinal structure function was then constructed from 
\begin{equation} 
F_L = {r^2 F_2 \over (1+R)}, 
\label{eq:fl1}
\end{equation}
with $r^2 = 1+{Q^2 \over {\nu}^2 } = {1 + {4 M^2 x^2 \over Q^2}}$.
For this purpose the phenomenological ALLM fit~\cite{allm97} was utilized, which 
reproduces the world $F_2$ data to better than 3\% and we include this as an additional uncertainty 
when propagating the uncertainties on $F_L$ ($F_1$), which are ultimately dominated by the 
uncertainties on the $R$ data.  The data sets utilized for the $F_L$ fit are listed in 
Table~\ref{datasets-fl}.

Before discussing the procedure, we feel it is worth commenting further on the need for 
uncorrelated data sets for $F_L$ and $F_2$.  This is also important for extractions of PDFs 
utilizing pQCD fits, since $F_L$ 
provides direct information on the gluon content that is only present in $F_2$ through the shallow 
logarithmic scaling violations.   In principle, the most information on the structure can be 
obtained from the full data set on the reduced cross section, which is proportional to 
$r^2F_2 - (1-\epsilon) F_L(x,Q^2)$, with $\epsilon$ a kinematic factor describing the photon 
polarization.  Most of the DIS data sets extract $F_2$ utilizing {\it all} the cross section 
measurements, after first determining $R$ by a linear fit in $\epsilon$ to the reduced cross 
section measurements at different $\epsilon$ values at fixed $x$ and $Q^2$ (the Rosenbluth 
separation technique~\cite{rosenbluth}).


\section{Procedure}

We follow the formalism for including TMCs to the massless limit structure function developed by 
Georgi and Politzer~\cite{gp} within the operator product expansion (OPE).  This approach was 
extended by Kretzer and Reno~\cite{kretzer_reno} to include the full set of 
electro-weak structure functions 
using modern formalism.  For a recent review of TMCs see~\cite{tmc-review}.  

At fixed $Q^2$ the $F_2$ structure function, including TM effects, can be written
\begin{equation}
F_2^{TM}(x,Q^2) =  {x^2 \over r^3} {F_2^{(0)}(\xi,Q^2) \over \xi^2}  
+ 6 {M^2 \over Q^2}  {x^3 \over r^4} \int_{\xi}^1 dx^{\prime}~{F_2^{(0)}(x^{\prime},Q^2) \over {x^{\prime}}^2} 
+ 12 {M^4 \over Q^4} {x^4 \over r^5} \int_{\xi}^1 dx^{\prime} \int_{x^{\prime}}^1 dx^{\prime \prime}~{F_2^{(0)}(x^{\prime \prime},Q^2) 
\over {x^{\prime \prime}}^2}
\label{eq:tmcf2}
\end{equation}
Here, $F_2^{(0)}(x,Q^2)$ is the massless limit structure function which can be calculated in pQCD 
utilizing the PDFs determined at some scale, $\xi~=~2x/(1+r)$ is the Nachtmann scaling variable, 
and both $x^{\prime}$ and $x^{\prime \prime}$ are dummy integration variables.  The corresponding TMC 
equation for $F_L$ is
\begin{equation}
F_L^{TM}(x,Q^2) =  {x^2 \over r} {F_L^{(0)}(\xi,Q^2) \over {\xi}^2}  
+ {4M^2 \over Q^2}  {x^3 \over r^2} \int_{\xi}^1 dx^{\prime}~{F_2^{(0)}(x^{\prime},Q^2) \over {x^{\prime}}^2} 
+ {8M^4 \over Q^4} {x^4 \over r^3} \int_{\xi}^1 dx^{\prime} \int_{x^{\prime}}^1 dx^{\prime \prime}~{F_2^{(0)}(x^{\prime \prime},Q^2) 
\over {x^{\prime \prime}}^2}.
\label{eq:tmcf1}
\end{equation}
The transverse structure function including the TM contributions, can then be constructed from
\begin{equation}
2xF_1^{TM} = {F_2^{TM} - F_L^{TM} \over {r^2}}  
\end{equation}
and in the massless limit ($r \rightarrow 0$) from
\begin{equation}
2xF_1^{(0)}  = F_2^{(0)} - F_L^{(0)}. 
\end{equation}

We note that these equations are valid to all orders in pQCD at twist 2. 
The target mass corrections for 
for higher twist operators in the unpolarized case are yet unknown. Moreover, dynamical higher twist 
contributions, starting with twist 4, cannot be described with one scaling variable only, 
neither do their potential target mass corrections, because the main variable $x$ is supplemented 
by further other invariants $x_i$. Their number grows with the level of twist included. This implies
the necessity to study different multi-parton correlations, which cannot be determined by inclusive 
measurements as that of the deep-inelastic structure functions only. Because of this 
the present analysis remains phenomenological applying the twist 2 target mass corrections to the
structure functions.~\footnote{In Ref.~\cite{Blumlein:1998nv} also the fermionic twist-3 target mass 
corrections in case of polarized deep-inelastic scattering have been calculated. They operate on 
the structure functions similar to the case of the twist-2 corrections as long as no gluonic operators 
in the respective region, i.e. at large $x$, are relevant. The polarized twist-3 TMCs are described
by different integrals as those at twist 2. However, quantitative numerical case studies using the 
same shape for input distributions have not been performed yet.}

While it is $F_2^{TM}(x,Q^2)$ that is measured in scattering experiments, it is the $Q^2$ evolution of 
$F_2^{(0)}$ that is described by pQCD.  To extract $F_2^{(0)}$ from data requires the inversion 
of Eq.~(\ref{eq:tmcf2}).  The procedure for extracting the TM corrected $F_2$ structure 
function is to parameterize $F_2^{(0)}(x,Q^2)$ and then to minimize the $\chi^2$ difference between 
the left-hand side of Eq~(\ref{eq:tmcf2}) and the $F_2$ data: 
\begin{equation}
\chi^2 =  \Sigma  \left ( {F_2^{TM}-F_2^{i} \over {\delta F_2^{i}}} \right )^2 ,
\label{chi2diff}
\end{equation}
where $F_2^{i}$ is the value of the $i^{th}$ data point and $\delta F_2^{i}$ it's uncorrelated 
uncertainty.

\begin{table}[tbh]
\caption{Data sets used in $F_2$ fit.}
\begin{center}
\begin{tabular}{l c c c c c}
\hline
\hline
Data Set  & $Q^2_{Min}$     &  $x_{min}$   &   $Q^2_{Max}$   &   $x_{max}$   &  \# Data Points      \\
          &  ($\rm GeV^2$)  &              &  ($\rm GeV^2$)  &               &                  \\  
\hline
BCDMS~\cite{f2rbcdms} &  7.5            &  0.070       &    230          &    0.75       &    178                 \\
NMC~\cite{f2rnmc}     &  0.75           &  0.0045      &    65           &    0.50       &    158                 \\  
SLAC (Whitlow~\cite{f2whitlow})   &  0.58    &  0.063  &    30           &    0.90       &    661                 \\
H1~\cite{f2h1e-,f2h1e+}        &   1.5   &  $3.0x10^{-5}$   & 5000    &   0.32    &  286   \\
Zeus~\cite{f2zeuse-,f2zeuse+}  &   3.5    &  $6.3 \times 10^{-5}$   & 5000     & 0.20   &  228   \\
E99-118~\cite{f2e99118}        &  0.273   &  0.077     &  1.67   &  0.320   &  9   \\
E94-110~\cite{e94110}          &  0.7   &  0.19    &   3.5   &  0.57   &    12 \\

\hline
\hline
\end{tabular}
\label{datasets-f2}
\end{center}
\end{table}
\begin{table}[tbh]
\caption{Data sets used in $F_L$ ($F_1$) fit.}
\begin{center}
\begin{tabular}{l c c c c c}
\hline
\hline
Data Set  & $Q^2_{Min}$     &  $x_{min}$   &   $Q^2_{Max}$   &   $x_{max}$   &  \# Data Points      \\
         &  ($\rm GeV^2$)  &              &  ($\rm GeV^2$)  &               &                    \\  
\hline
BCDMS~\cite{f2rbcdms}            &  15      &  0.07      &  50     &  0.65    & 10   \\
EMC~\cite{f2remc}                &  15      &  0.041     &  90     &  0.369   & 28   \\
NMC~\cite{f2rnmc}              &  1.31    &  0.0045    &  20.6   &  0.11    & 10   \\  
SLAC (Whitlow~\cite{rwhitlow}) &  0.63    &  0.1       &  20     &  0.86    & 90   \\
SLAC (E140x~\cite{re140x})     &  0.5     &  0.1       &   3.6   &  0.50    &  4   \\
H1~\cite{flh1}                 &  1.5      &  $3.0 \times 10^{-5}$  &  45     &  0.0015  &  13   \\
ZEUS~\cite{flzeus}             &  24      &  $6.7 \times 10^{-4}$   & 110  & 0.0049 & 18  \\   
E99-118~\cite{re99118}         &  0.273   &  0.077     &  1.67   &  0.320   &  8   \\
E94-110~\cite{e94110}          &  0.7     &  0.19      & 3.5     &  0.57    & 12   \\
\hline
\hline
\end{tabular}
\label{datasets-fl}
\end{center}
\end{table}

The fit form utilized for the $x$ dependence of both $F_2^{(0)}$ and $F_L^{(0)}$ was
\begin{equation}
F_{2,L}^{(0)}(x) = A x^B (1-x)^C (1+D \sqrt x + E x),
\end{equation}
with the $Q^2$ dependence of A, B , C, D, and E parameterized as
\begin{equation}
A(Q^2) = A_1 + A_2 e^{-Q^2/A_3} + A_4 \log (0.3^2+Q^2).
\end{equation}
These forms are empirical, but were found to give a good description of the data.
The fitting was performed utilizing {\tt MINUIT}~\cite{minuit} and utilized an
optimized {\tt FORTRAN}
routine {\tt DAIND}~\cite{daind} to perform the single and double integrations in 
Eqs.~(\ref{eq:tmcf2}) and~(\ref{eq:tmcf1}).

\section{Results and Conclusions}

\begin{table}[tbh]
\caption{$F_2$ best fit parameters.}
\begin{center}
\begin{tabular}{l c c c c}
\hline
\hline
   $n = $ &     1        &     2        &     3           &   4              \\ 
\hline
 $A_n^2$  &   0.2145       & 0       & 0    &  0  \\
 $B_n^2$  &   -2.1540       & .0808    &    5.6730   &  -0.0319  \\
 $C_n^2$  &    2.7830       & -0.5013   &    7.104       &   0.1857    \\
 $D_n^2$  &  -0.8997       &  -3.1092  &   0.3820 &  -0.1275       \\
 $E_n^2$  &    9.0441      & 0       & 0   &  0  \\
\hline
\hline
\end{tabular}
\label{param_f2}
\end{center}
\end{table}
\begin{table}[tbh]
\caption{$F_L$ best fit parameters. Parameter $C_1^L$ (indicated by *) has been fixed, 
as the goodness of the fit was found to be only minimally sensitive to it's value.}
\begin{center}
\begin{tabular}{l c c c c}
\hline
\hline
   $n = $ &     1        &     2        &     3           &   4              \\ 
\hline
 $A_n^L$  &    0.0078        & 0.0 & 0.0 &  0.0  \\
 $B_n^L$  &   -2.0650        & 0.0 & 0.0 &   -4.8846                   \\
 $C_n^L$  &    1.0*          & 0.0 & 0.0 &    0.0                      \\
 $D_n^L$  &    0.0           &  -0.0030 &   0.0  &   0.6040       \\
 $E_n^L$  &    0.0           & 0.0       & 0.0 &  0.0  \\
\hline
\hline
\end{tabular}
\label{param_fl}
\end{center}
\end{table}

%

\begin{figure}
\includegraphics[width=8.8cm,height=13cm]{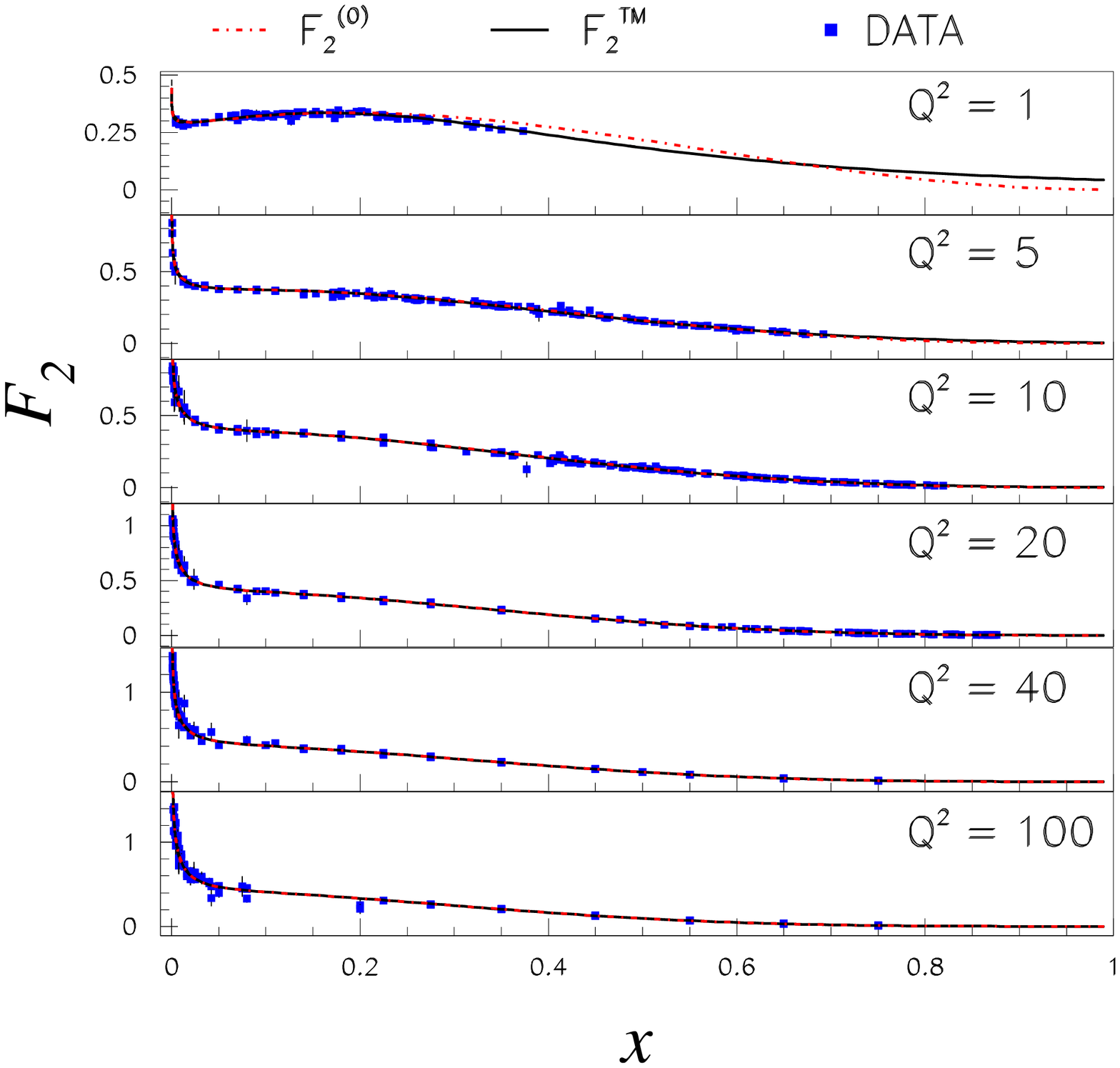}
\includegraphics[width=8.8cm,height=13cm]{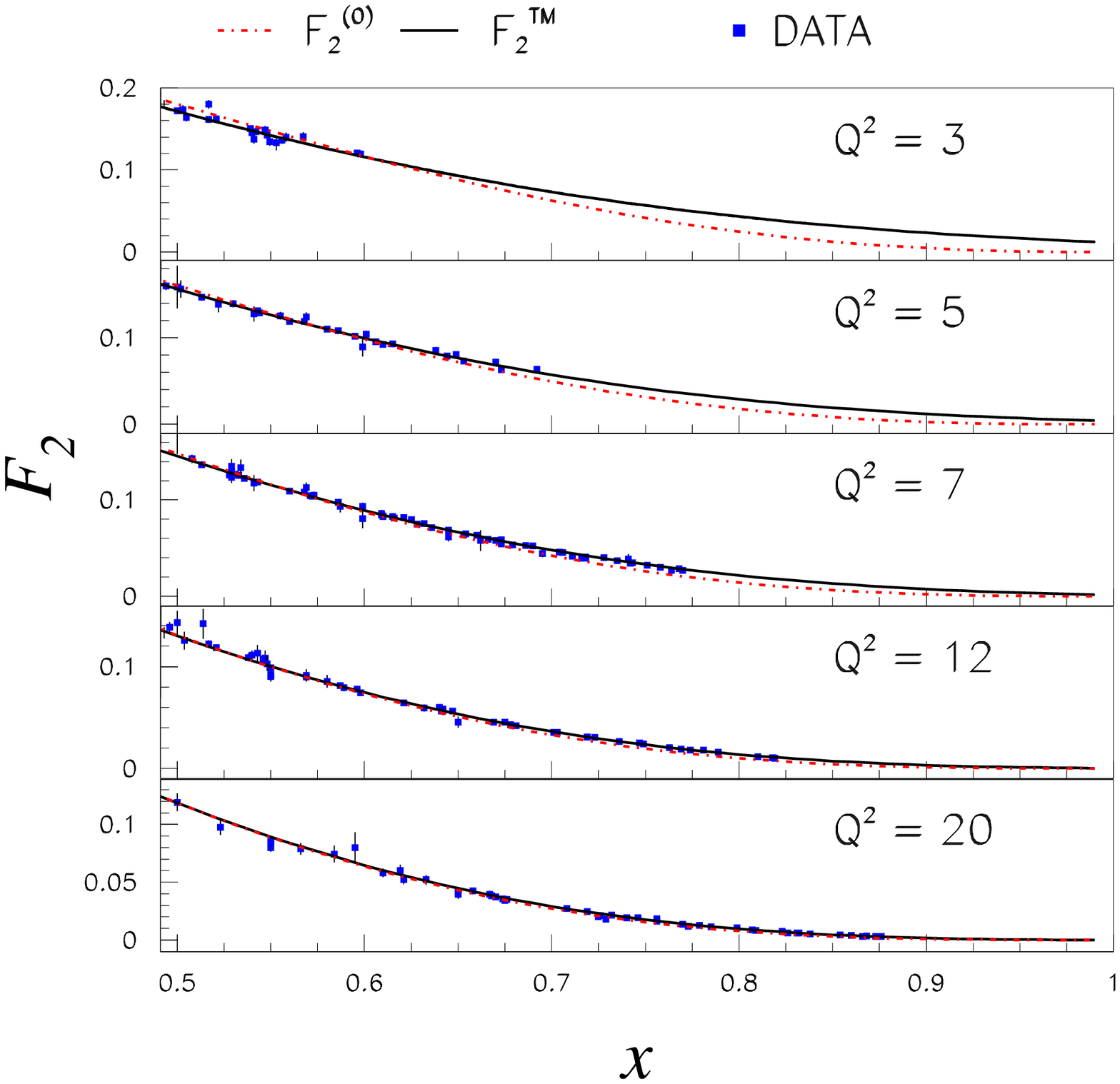}
\caption{\label{f2_panel} (Color online) Comparison of the $F_2^{TM}$ fit results 
(dashed curve) to the proton data for 6 different $Q^2$ values (left panel). The low $x$ data from 
H1 and ZEUS are beyond the vertical scale.
Also shown is the massless limit structure function parameterization from the fit (solid curve).  A
zoom view for the region $x~>~0.5$ is also shown for 5 different $Q^2$ values (right panel).}
\label{fig:f2results}
\end{figure}

\begin{figure}
\includegraphics[width=8.8cm,height=11cm]{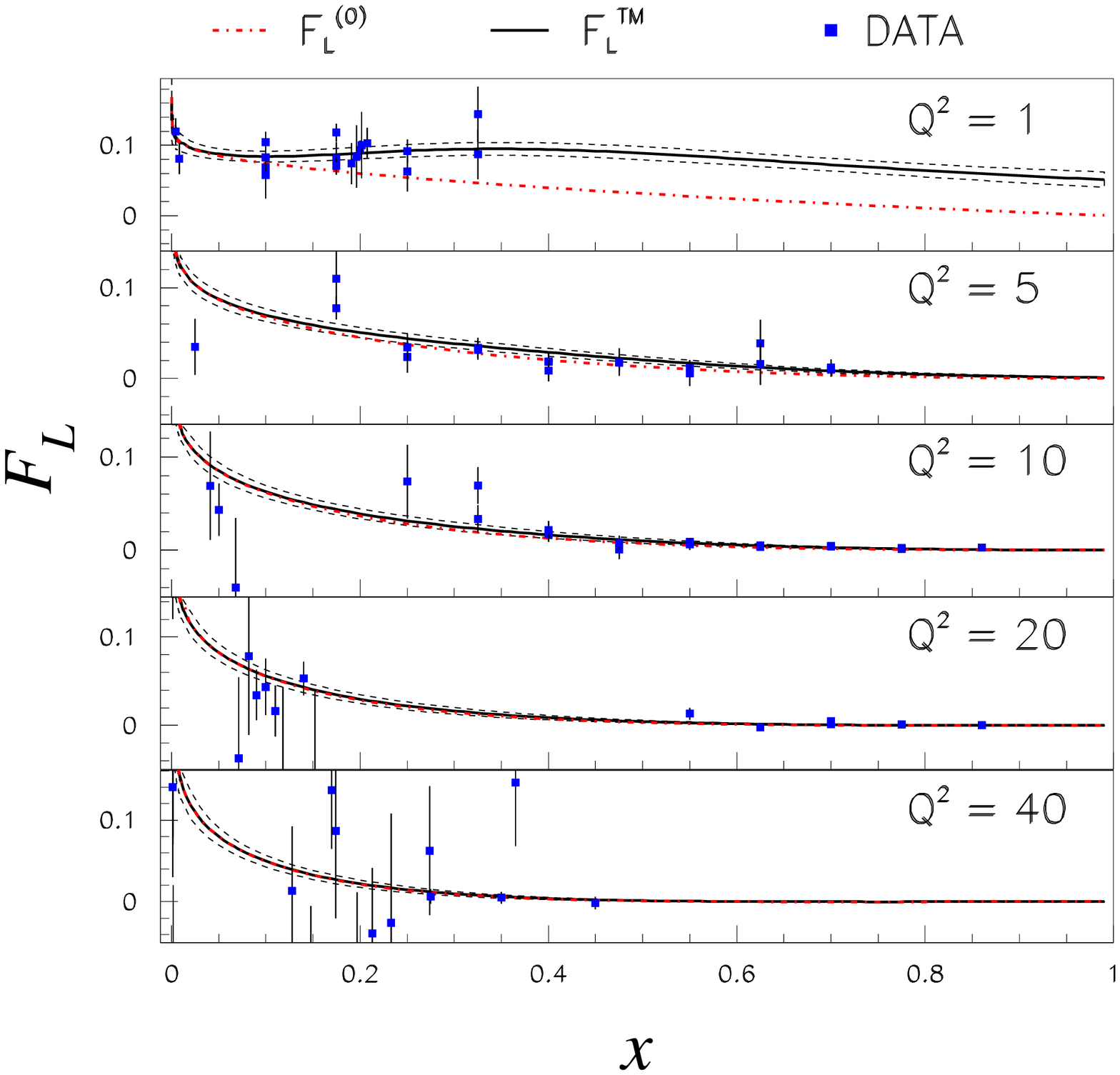}
\includegraphics[width=8.8cm,height=11cm]{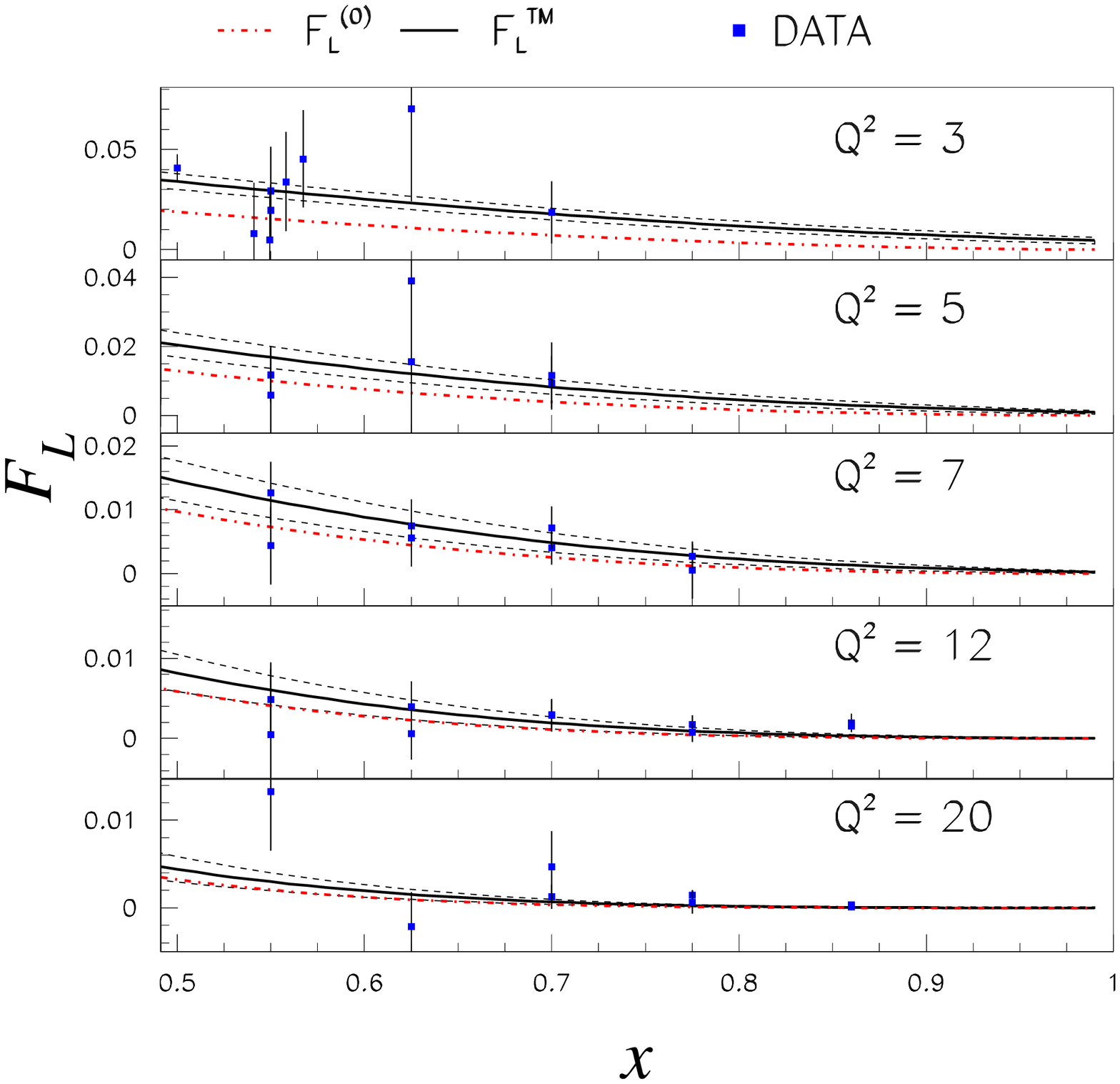}
\caption{\label{fig:flresults} 
(Color online) Comparison of the $F_L^{TM}$ fit results 
(solid curve) to the proton data versus $x$ for 5 different $Q^2$ values (left panel).  The low $x$ 
data from H1 and ZEUS are beyond the vertical scale. 
Also shown is the massless limit structure function parameterization from the fit (dot-dashed curve).     
A zoom view for the region $x~>~0.5$ is also shown (right panel).  
All data within a range of $\pm 40\%$ ($Q^2=1$), $\pm 33\%$ ($Q^2=3$), 
$\pm 15\%$ ($Q^2=5$, 7, 10, and 12), and $\pm 20\%$ ($Q^2=20$ and 40) have been bin-centered 
to the central $Q^2$ using the $Q^2$ dependence from the fit. 
The uncertainty band stemming from the uncorrelated uncertainties and calculated from the fit 
covariance matrix is indicated by the dashed curve.}
\end{figure}

The fit results for both $F_2$ and $F_L$ are quite good, with $\chi^2$ per degree of freedom 
($\chi^2$/DoF) of 1.01 and 0.93 for $F_2$ and $F_L$, respectively.  The resulting fit parameters 
are listed in Table~\ref{param_f2} for $F_2$ and Table~\ref{param_fl} for $F_L$.  Comparisons of 
the fit results to the $F_2$ data are shown in Figure~\ref{fig:f2results} for the 6 different 
$Q^2$ values listed on the panels from 1 to 100~$\rm GeV^2$.  Data within a range of $Q^2$ about the 
central value have been centered to the common $Q^2$ value using the fit.  A comparison over the 
full $x$ range fit is shown (Left), as well as for large $x$ (Right). 
Similar comparisons for $F_L$ are shown in Figure~\ref{fig:flresults}.  Of particular note is the 
lack of quality data for $F_L$ in the range $0.005 < x <0.3$ for $Q^2 > 3$~$\rm GeV^2$.  Such data 
is critical for constraining the fit at low $x$ and would be of particular value in constraining 
the gluon distribution which is a dominant contribution to $F_L$. 

We expect that the fit will be of interest to the broader community in providing a good description 
of the data on the full set of unpolarized structure functions in charged lepton scattering from the 
proton, $F_2$, $F_L$, $F_1$, and $R$, over the kinematic range of $x > 3 \times 10^{-5}$ and 
$0.1 <Q^2~\rm GeV^2$, in addition to providing a separation of the TMCs from the massless limit 
structure functions.  

The separated 
structure functions can be used for further pQCD studies, such as HT analyses of both the structure 
functions and moments.  The current analysis procedure is complementary to PDF fits which include 
TMCs in that the unfolding of the TMCs does not depend on a finite order expansion in $\alpha_S$.  
Comparison of the results from the two procedures could provide an additional handle on 
isolating HT contributions from higher-order QCD corrections and making for a 'cleaner' evaluation 
of the former.

Additionally, the fit will be of use in future studies of quark-hadron duality.  In particular, 
determining how well the {\it averaged} resonance region structure functions at fixed $x$ obey 
the $Q^2$ dependence dictated from DIS.    

{\tt FORTRAN} computer code embodying the fit described in this article is available by email request 
from the authors, including the full covariance matrices.

\begin{acknowledgments}
We would like to thank R. Ent, C.E. Keppel, and A. Accardi for useful discussions.  
This work was supported in part by research grant 1002644 from the National Science Foundation and
by the Deutsche Forschungsgemeinschaft in Sonderforschungs\-be\-reich/Transregio~9
and by the European Commission through contract PITN-GA-2010-264564 ({LHCPhenoNet}).
 
\end{acknowledgments}



\end{document}